\documentclass[11pt]{article}

\begin{document}
\title{Transport Properties Calculation for a Quasi-Bidimensional System\\
 using T-Matrix Approximation }
\author{C. P. Moca\footnote{Corresponding author. Fax +40-59-432789;E-mail:mocap@ff.uoradea.ro}\hspace{0.15cm} and E. Macocian \\
Department of Physics, University of Oradea\\
3700, Oradea, Romania}
\maketitle

\begin{abstract}
We performed a self-consistent calculation using T-Matrix approximation for
a quasi-bidimensional system. We calculated the one particle spectrum
function $A\left( \mathbf{k},\omega\right) $ in the presence of strong
d-wave attractive interaction. The $c$-axis charge dynamics was studied by
considering incoherent interlayer hopping and $ab$-plane charge dynamics was
studied in the coherent limit. It is shown that the $c$-axis charge dynamics
is mainly governed by the scattering from the in plane fluctuations. We also
present results for $c$-axis and $ab$-plane resistivity and for thermopower
coefficient. 
\end{abstract}

\section{Introduction}

In recent years one of the significant advances in the research of high-T$_c$
superconductors is the detailed experiments on the pseudogap behavior in
underdoped high-T$_c$ superconductors. The pseudogap phenomena have been
recognized as one of the most important issues. There are enormous studies
from both experimental and theoretical point of view. Influence of the
pseudogap formation shows up in different experimental probes, such as NMR
relaxation time, Knight shift, neutron scattering, tunneling, photoemission,
specific heat, optical conductivity and d.c. resistivity\cite{1}.

The pseudogap phenomena means the suppression of the low frequency spectral
weight without any long range order. For cuprate superconductors the low
frequency spectral weight begins to be strongly suppressed below some
characteristic temperature T$^{*}$ much higher than T$_c$. Typically T$^{*}$
is several times larger that T$_c$ in underdoped cuprates and their doping
dependence is qualitatively different. While T$_c$ deceases with
underdoping, T$^{*}$ increases in contrast. This suggest that T$^{*}$ does
not follow the real T$_c$ line but instead some kind of mean-field critical
temperature T$_{MF}$.

In reality the origin of pairing interaction should be considered to be the
antiferromagnetic spin fluctuations. There are studies dealing with pairing
correlations obtained by the spin fluctuations on the basis of the
fluctuation-exchange (FLEX) approximation \cite{2}. To explain the pseudogap
phenomena many theories have been proposed, such as the argument based on
BCS-BE cross-over \cite{3}, spin-charge separation induced by stripe order 
\cite{4}, RVB theory \cite{5}, strong superconducting phase fluctuations 
\cite{6,7}, magnetic scenario near the antiferromagnetic instability \cite
{8,9,10} and formation of charge density waves (CDW) or spin density waves
(SDW) \cite{49}. A feature of the last scenario is that the pseudogap can
persist in the superconducting state \cite{50}. The effect of the magnetic
field on the pseudogap in high-T$_c$ cuprates has been theoretically
investigated by Yanase et.al. \cite{11} considering the Landau quantization
for the superconducting fluctuations as the main effect of the magnetic
field. The pseudogap phenomena was studied in the context of non-Fermi
Anderson model in \cite{12} considering the electron-fluctuation
interaction. Recent dimensional crossover study by Preosti et.al. \cite{13}
shows that the pseudogap effect is basically absent in three dimensions.

In this paper we describe the pairing scenario based on the strong coupling
superconductivity. Generally the strong coupling superconductivity indicates
the existence of the incoherent Cooper pairs as the Nozieres and
Schmitt-Rink formalism \cite{14}. We think of the pseudogap as the phenomena
brought about by the strong superconducting fluctuations. Maly et.al. \cite
{15} have introduced the idea of ''resonances'' which means the scattering
mechanism by the presence of metastable preformed pairs.

Through several years of extensive experimental work general consensus
regarding the superconducting gap symmetry seems to be reach, that the
superconducting gap has mainly $d$-wave character with possibility of a
small mixture of other angular momentum state \cite{16,17,18}. A remarkable
point of the pseudogap is that its structure in momentum space is the same
as the superconducting $d_{x^2-y^2}$ symmetry with continuous evolution
through T$_c$. This implies that the pseudogap phenomena has close
connection to the superconducting fluctuations \cite{19}. On the other hand
using Intrinsic Tunneling Spectroscopy, Krasnov et.al. \cite{48} found that
the pseudogap is coexisting with the superconducting gap which imply that
the two phenomena are different in nature. The precursor superconductivity
scenario can not explain these experimental data and a possible explication
might be the formation of CDW or SDW \cite{49}. Another striking
characteristic of cuprate superconductors is the anisotropy of their
structure. The quantity which most evidently displays the anisotropy of a
particular material is the optical conductivity $\sigma \left( \omega
,T\right) $ and the corresponding d.c. resistivity $\rho \left( T\right) $ 
\cite{20}.

The opening of the pseudogap has drastic effect on the physical properties
of the high-T$_c$ cuprates \cite{1}. It is found that associated with the
pseudogap the in-plane resistivity deviates from the $T$-linear behavior 
\cite{21} and the T coefficient of the $c$-axis resistivity changes sign,
showing semiconducting behavior at low temperature \cite{22}.

This paper is constructed as follow. In section 2 we give the model
Hamiltonian and explain the theoretical framework. In section 3 we present
the results obtained for the spectral function $A\left( \mathbf{k},\omega
\right) $ and the results for the transport properties studied in this
paper. In section 4 we summarize the obtained results and give conclusions.

\section{Theoretical Framework}

\subsection{Model Hamiltonian}

\ In this section we describe the theoretical framework of this paper. We
consider a microscopic model which incorporates both strong
electron-fluctuation interaction in the $CuO$ planes and a weak interlayer
coupling. Within each layer we consider the following two-dimensional model
Hamiltonian which has $d_{x^2-y^2}$-wave superconducting ground state: 
\begin{equation}
H_l=\sum\limits_{\mathbf{k},\sigma }\varepsilon _{\mathbf{k}}c_{l\mathbf{k}%
\sigma }^{+}c_{l\mathbf{k}\sigma }+\sum\limits_{\mathbf{q,k,k}^{\prime }}V_{%
\mathbf{kk}^{\prime }}c_{l\mathbf{k}\uparrow }^{+}c_{l\mathbf{k+q\uparrow }%
}c_{l\mathbf{k}^{\prime }\downarrow }^{+}c_{l\mathbf{k}^{\prime }-\mathbf{%
q\downarrow }}  \label{1}
\end{equation}
where $c_{l\mathbf{k}\sigma }$ is the annihilation operator for an electron
with momentum $\mathbf{k}$ and spin $\sigma $ in the layer $l$. The electron
dispersion relation is expressed as: 
\begin{equation}
\varepsilon _{\mathbf{k}}=-2t\left( \cos k_x+\cos k_y\right) -4t^{\prime
}\cos k_x\cos k_y-\mu  \label{2}
\end{equation}
where $t$ and $t^{\prime }$ are the nearest and next-nearest neighbors
hopping amplitudes and $\mu $ is the chemical potential. From now on we will
consider $t$ equal to unity and $t^{\prime }=-0.5t$. The pair interaction is
written as: 
\begin{equation}
V_{\mathbf{kk}^{\prime }}=Vf_{\mathbf{k}}f_{\mathbf{k}^{\prime }}  \label{3}
\end{equation}
where 
\begin{equation}
f_{\mathbf{k}}=\cos k_x-\cos k_y  \label{4}
\end{equation}
is the $d_{x^2-y^2}$-wave form factor. In the previous relation V is
negative. The hopping between layers is included in: 
\begin{equation}
H=\sum\limits_lH_l-t_c\sum\limits_l\left( c_{li\sigma }^{+}c_{l+1i\sigma
}+c_{l+1i\sigma }^{+}c_{li\sigma }\right)  \label{5}
\end{equation}
where $t_c$ is the hopping amplitude between layers and $c_{li\sigma }$ is
the annihilation operator for an electron within the planar site $i$, with
spin $\sigma $ in the layer $l$. The important character of high-T$_c$
cuprates is the momentum dependence of the interlayer hopping matrix element 
$t_c\left( \mathbf{k}\right) $. The band calculations \cite{25,26} of these
materials have shown that the c-axis tunneling matrix element is larger at $%
\left( \pi ,0\right) $ and symmetry related points. The functional form is: 
\begin{equation}
t_c\left( \mathbf{k}\right) =\left( \cos k_x-\cos k_y\right) ^2  \label{6}
\end{equation}
On the other side the coupling between layers has little effect on the
pseudogap \cite{27}. The out of plane resistance is determined mainly by the
properties of individuals $CuO$ layers. The large magnitude of the
resistivity anisotropy reflects that the $c$-axis mean free path is shorter
than the interlayer distance, and the carriers are tightly confined to the $%
CuO$ planes and also is the evidence of the incoherent charge dynamics in
the $c$-axis direction.

The interlayer part of the Hamiltonian (\ref{5}) will be considered for the
calculation of $c$-axis charge dynamics and $\rho _c\left( T\right) $.

\subsection{Self-Consistent T-Matrix Approximation}

In this paper we focus on the calculation of the Green function directly on
the real-frequency axis \cite{23} in order to avoid the difficulties of
controlling the accuracy of the calculations used in the numerical
analytical continuations from the imaginary to real axis by Pade algorithm 
\cite{24}.

We carry out a self-consistent calculation for the spectral function: 
\begin{equation}
A\left( \mathbf{k},\omega \right) =-\frac 1\pi ImG\left( \mathbf{k},\omega
\right)   \label{7}
\end{equation}
where the retarded Green function $G\left( \mathbf{k},\omega \right) $ is
given by: 
\begin{equation}
G\left( \mathbf{k},\omega \right) =\frac 1{\omega -\varepsilon _{\mathbf{k}%
}-\Sigma \left( \mathbf{k},\omega \right) }  \label{8}
\end{equation}
where $\Sigma \left( \mathbf{k},\omega \right) $ is the retarded
self-energy. The imaginary part of the self-energy can be expressed as: 
\begin{equation}
Im\Sigma \left( \mathbf{k},\omega \right) =f_{\mathbf{k}}^2\sum\limits_{%
\mathbf{k}^{\prime }}\int d\omega ^{\prime }\left[ f\left( \omega ^{\prime
}\right) +n\left( \omega +\omega ^{\prime }\right) \right] A\left( \mathbf{k}%
^{\prime },\omega ^{\prime }\right) ImT\left( \mathbf{k}+\mathbf{k}^{\prime
},\omega +\omega ^{\prime }\right)   \label{9}
\end{equation}
where $f\left( \omega \right) $ and $n\left( \omega \right) $ represent
Fermi-Dirac and Bose-Einstein distributions. The real part of the retarded
self-energy can be calculated directly or using Kramers-Kronig relation: 
\begin{equation}
Re\Sigma \left( \mathbf{k},\omega \right) =p.v.\int \frac{d\omega ^{\prime }}%
\pi \frac{Im\Sigma \left( \mathbf{k},\omega ^{\prime }\right) }{\omega
-\omega ^{\prime }}  \label{10}
\end{equation}
where $p.v.$ represents the principal value of the integral. The Hartree
term is neglected in the self-energy. The T-Matrix is given by the following
relation: 
\begin{equation}
T\left( \mathbf{k},\omega \right) =\frac{V^2\Pi \left( \mathbf{k},\omega
\right) }{1+V\Pi \left( \mathbf{k},\omega \right) }  \label{11}
\end{equation}
where $\Pi \left( \mathbf{k},\omega \right) $ can be evaluated directly from
the spectral function as: 
\begin{equation}
Im\Pi \left( \mathbf{q},\omega \right) =\pi \sum\limits_{\mathbf{k}}\int
d\omega ^{\prime }f_{\mathbf{k}}^2\tanh \frac \omega {2T}A\left( \mathbf{k}%
,\omega ^{\prime }\right) A\left( \mathbf{q}-\mathbf{k},\omega -\omega
^{\prime }\right)   \label{12}
\end{equation}
The chemical potential is determined by fixing the density $n$ through the
relation: 
\begin{equation}
\frac n2=\sum\limits_{\mathbf{k}}\int d\omega A\left( \mathbf{k},\omega
\right) f\left( \omega \right)   \label{13}
\end{equation}
We can calculate the hole doping $\delta $ using the relation $\delta =1-n$.
Beside relation (\ref{13}) we must take care of the sum-rule satisfied by
the spectral function: 
\begin{equation}
1=\sum\limits_{\mathbf{k}}\int d\omega A\left( \mathbf{k},\omega \right) 
\label{14}
\end{equation}
which must be evaluated at each step in the self-consistent calculation. In
the present calculation we divided first Brilouinne zone into $32\times 32$
or $64\times 64$ lattice and the frequency integration was replaced by a
discrete sum of $1024$ or $512$ points. We did not found differences in the
behavior of spectral function using these two discretizations.

Performing the self-consistent calculation given by Eqs. (\ref{7})-(\ref{14}%
) we obtain the spectral function $A\left( \mathbf{k},\omega \right) $. We
also calculated the density of states: 
\begin{equation}
N\left( \omega \right) =\sum\limits_{\mathbf{k}}A\left( \mathbf{k},\omega
\right)  \label{15}
\end{equation}
Having obtained the spectral function we can calculate different physical
properties of the cuprates. The real part of the $ab$-plane optical
conductivity is given by: 
\begin{equation}
\sigma _{ab}\left( \nu \right) =\sigma _0\int d\omega \frac{f\left( \omega
\right) -f\left( \omega +\nu \right) }\nu \sum\limits_{\mathbf{k}}\left[
\left( \frac{\partial \varepsilon _{\mathbf{k}}}{\partial k_x}\right)
^2+\left( \frac{\partial \varepsilon _{\mathbf{k}}}{\partial k_y}\right)
^2\right] A\left( \mathbf{k},\omega \right) A\left( \mathbf{k},\omega +\nu
\right)  \label{16}
\end{equation}
where $\sigma _0=e^2\pi /2h$.

The in-plane conductivity $\sigma _{ab}\left( \omega \right) $ given by eq. (%
\ref{16}) neglects the vertex correction. The in-plane conductivity is
coherent in character and shows a Drude peak at low frequencies.

For the calculation of $c$-axis conductivity $\sigma _c\left( \omega \right) 
$ we consider the incoherent case. Incoherent conductivity corresponds to
diffusive $c$-axis transmission and amounts in taking the averages of the
spectral function $N\left( \omega \right) $ over all momenta. 
\begin{equation}
\sigma _c\left( \nu \right) =\sigma _0\int d\omega \frac{f\left( \omega
\right) -f\left( \omega +\nu \right) }\nu \sum\limits_{\mathbf{k}%
}t_c^2\left( \mathbf{k}\right) N\left( \omega \right) N\left( \omega +\nu
\right)  \label{17}
\end{equation}
The same relation appear in the calculation of optical conductivity in
disordered systems \cite{28}. We calculate in the limit $\omega \rightarrow
0 $ the in-plane resistivity $\rho _{ab}=\lim_{\omega \rightarrow 0} (1/{%
\sigma _{ab}\left( \omega \right) })$ and $c$-axis resistivity $\rho
_c=\lim_{\omega \rightarrow 0} (1/{\sigma _c\left( \omega \right) })$.

In order to evaluate the thermopower we have to calculate \cite{29}: 
\begin{equation}
L_{ij}=\int d\omega \left( -\frac{\partial f\left( \omega \right) }{\partial
\omega }\right) \left[ \sum\limits_{\mathbf{k}}\left( \frac{\partial
\varepsilon _{\mathbf{k}}}{\partial k_x}\right) ^2A^2\left( \mathbf{k}%
,\omega \right) \right] ^i\omega ^{j-1}  \label{18}
\end{equation}
and the thermopower can be calculated using: 
\begin{equation}
S=-\frac{k_B}{\left| e\right| T}\frac{L_{12}}{L_{11}}  \label{19}
\end{equation}

\section{Results}

\subsection{Spectral Function and Phase Diagram}

In this section we present the obtained results for spectral function, using
the self-consistent T-Matrix approximation described in previous section. In
order to make the results more transparent we present the lowest order
calculation results and the self-consistent calculations. We also present
the phase diagram obtained for this model. As we can see both in Fig. 1 and
Fig. 2 in the lowest order calculation we obtain a pseudogap along the
direction $\left(0,\pi \right) -\left( \pi ,\pi \right) $ and no pseudogap
is obtained along the $\left( 0,0\right) -\left( \pi ,\pi \right) $
direction. This is in agreement with ARPES measurements \cite{30}. Janko
et.al. claims that this anomaly is caused by nearly stable cooper pairs in
the excited states \cite{31}. 
The same behavior is observed in the self-consistent calculation as we can
see in Fig. 3 and Fig. 4. Qualitatively we found small differences in
behavior between lowest and self-consistent calculation for the spectral
function.

We also calculated the critical temperature T$_c$ of the system using
Thouless criterion \cite{14}. When $1+V\Pi \left( \mathbf{k},\omega \right)
=0$ the superconductivity occurs. Thouless criterion is equivalent to BCS
theory in the weak coupling limit. In our approximation $T(\mathbf{k},\omega
)$ can be regarded as the propagator of the Cooper pairs. Thouless criterion
correspond to the situation in which $T(\mathbf{k},\omega )$ has its pole at 
$\mathbf{k}=0$ and $\omega =0$. We are interested in the behavior of the
system in the normal state, near the superconducting critical point, where
the superconducting fluctuations are inhaced. 

In Fig. 5 we present the phase
diagram of the model where $T_c$ and $T_{MF}$ are the critical temperature
calculate in the self-consistent and lowest-order calculated using Thouless
criterion. We also calculated the critical temperature in the first order of
perturbation theory and found that in the limit of strong coupling the
self-consistent calculation of $T_C$ reach the limit of first order
calculation of $T_C$. $T_{MF}$ can be viewed as the BCS critical
temperature. The phase diagram corresponds to the underdoped region with the
hole concentration $\delta =0.1$. Roughly speaking the pseudogap phenomena
occurs in the region between T$_c$ and T$_{MF}$. This region is extremely
wide in the strong coupling case. As we can see from the shape of spectral
function $A\left( \mathbf{k},\omega \right) $ the main effect of the
electron-fluctuation interaction is the change of shape of $A\left( \mathbf{k%
},\omega \right) $ which is seen near $\mathbf{k}=\left( 0,\pi \right) $.
This suggests that the electron-fluctuation interaction shifts the Fermi
surface to the place far from $\mathbf{k}=\left( 0,\pi \right) $. 
Similar with other reported self-consistent treatments of the T-matrix \cite
{51,52} the pseudogap is not seen clearly in the integrated density of
states but only in the $\mathbf{k}$-dependent spectral function at extreme
parameters ranges. In Fig. 6a we present the density of states for different
couplings at the temperature $T=0.35t$ and in Fig. 6b the temperature
dependence of the density of states for fixed coupling strength $V=-3t$. As
we can see from Fig. 6a, increasing the coupling strength, the density of
states near $\omega =0$ slowly decreases but the central peak still remain.
The central peak is due to the integration over the direction $(0,0)-(\pi
,\pi )$ in the spectral function where no pseudogap appears. For the same
temperature the central peak remain at the same frequency $\omega \approx 0$%
. Decreasing the temperature the central peak start to shifts to lower
frequencies and a hump start to appear in the density of states. The two
humps appearing in Fig. 6a are due to the pseudogap formation in the
spectral function near the $(0,\pi )$ and symmetry related points and their
intensity increase with lowering the temperature. 

\subsection{Optical Conductivity and Resistivity}

We have calculated the in-plane and $c$-axis optical conductivity using eqs.
(\ref{16}) and (\ref{17}). We also have calculated the resistivity as
function of temperature for both in-plane and $c$-axis direction. Our
approximation assumes the independent electron propagation in each layer and
is justified for small interlayer hopping. The interlayer hopping hopping
matrix element $t_c$ is weak resulting in an incoherent c-axis transport.
Many approaches have been suggested \cite{32,33,34} to describe the c-axis
transport properties. The optical conductivity was calculated combining well
controlled analytical and numerical methods in the case of one-dimensional
Mott-Hubbard insulator at zero temperature in \cite{35}. The $c$-axis
optical conductivity and d.c. resistivity was calculated within $t-J$ model
in \cite{39}.

In Fig. 7 we present the obtained results for the in-plane conductivity as
function of frequency for two temperatures. The in-plane spectrum of $\sigma
_{ab}\left( \omega \right) $ is dominated by Drude peak at $\omega =0$. The
intensity of the peak grows by increasing the temperature.
In Fig. 8 we present the $c$-axis optical conductivity for three
temperatures at the doping level $\delta =0.1$. The electronic contribution
to $\sigma _c\left( \omega \right) $ in the normal state of the underdoped
cuprates is nearly $\omega $ independent and does nor form a Drude peak at $%
\omega =0$, indicating incoherent nature of the charge dynamics along the $c$%
-axis \cite{36}. It should be emphasize that the pseudogap appears only in $%
\sigma _c\left( \omega \right) $. Actually the development of the pseudogap
in $\sigma _c\left( \omega \right) $ is correlated with a rapid narrowing of
Drude width in $\sigma _c\left( \omega \right) $. We observe an anomalous
broad peak in $\sigma _c\left( \omega \right) $ at finite frequencies which
grows in intensity as the temperature is lowered with no discontinuity at
the superconducting transition similar to the experimental results \cite{37}%
. In Fig. 9 we present the temperature dependence of the in-plane resistivity $%
\rho _{ab}$. We found a linear behavior for the in-plane resistivity at high
temperature and a deviation from linearity below a characteristic
temperature $T^{*}.$ $T^{*}$ may be defined by different methods \cite{53}.
In this paper we consider the temperature $T^{*}$ the characteristic
temperature where the resistivity present a downturn, with a linear behavior
above it. The arrows in Fig. 9 indicate $T^{*}$. Increasing the doping
concentration $\delta $, $T^{*}$ decrease. 

In Fig. 10 we present the $c$-axis resistivity. We found in this case the
same linear behavior of resistivity as function of temperatures, at high
temperatures and a upturn at lower temperatures $\left( d\rho _c/dT<0\right) 
$ signaling a semiconductor -like behavior. 
The semiconducting $T$ dependence of the $c$-axis resistivity is associated
with the deepening of the pseudogap. Therefore the contribution to the
conductivity from the quasiparticles near the $\left( \pi ,0\right) $ is
suppressed by the pseudogap. The in-plane conductivity is determined by the
contributions from ''cold-spots'' and the pseudogap is not important for the
in-plane conductivity, but has more effect on the $c$-axis conductivity. The 
$c$-axis conductivity is determined from the contributions from the
''hot-spots'' where the pseudogap is large. Therefore the $c$-axis
conductivity is remarkably suppressed by the pseudogap. This is in agreement
with the experimental results for cuprates \cite{38}. Our results are
qualitatively in agreement with the experimental results obtained for $%
\left( R_{1-x}Ca_x\right) Ba_2Cu_3O_{6+\delta }$ where with $Ca$ doping the
number of holes in $CuO_2$ planes increases \cite{41,42}, where $R$ stands
for $Y$ and $Lu$. $\rho _{ab}$ is linear in $T$ above $T^{*}$ and decreases
more steeply below $T^{*}$ \cite{21}.

\subsection{Thermopower}

Thermopower has become a simple and ideal probe for the studies of high $T_C$
superconductors. The thermopower is one of the several physical quantities
which distinctly reveals the unusual normal state properties. The results
obtained for the thermopower are calculated using eq. (\ref{19}). The
thermopower of cuprates \cite{40} has the propriety that is not a monotonic
function of temperature. Several authors have tried to explain the
temperature behavior of thermopower in term of phonon drag, metallic
diffusion, semiconducting model and others \cite{55}, but they cannot
explain the linear behavior at high temperature \cite{46} without
considering the cylindrical Fermi surface \cite{47}. 

In Fig. 11 we present the temperature dependence of the thermopower. It can be seen that the
normal state thermopower decrease with increasing hole doping $\delta \,$ to
a lower positive value, at a fixed temperature. The temperature dependence
of thermopower is linear with a negative slope at high temperature and
present a broad peak, shifted toward the low-temperature as the doping
increase \cite{43}. The hump of thermopower in the normal state was reported
as a kind of universal behavior of $HTSC$, corelated with the opening of the
pseudogap in the normal state of a superconductor \cite{54}. 

In Fig. 12 we present the doping dependence of the thermopower at the
interaction strength $V=-5.5t$ and at the temperature $T=2t$. At a fixed
temperature much larger than the critical temperature, $S$ continuously
decreases from underdoped to overdoped regions, changing its sign near
optimal doping. The theoretical work presented in \cite{45} shows that the
doping dependence of the thermopower can be explained by the common band
dispersion relation of high-T$_c$ cuprates. 
Therefore the thermopower can be a criterium for measuring the hole doping
in $CuO$ layers \cite{44}.

\section{Conclusions}

In this paper we have shown that the d-wave pairing scenario, based on the
strong coupling superconductivity well explains some of the physical
properties of the layered cuprates above the critical temperature $T_c$. We
have considered that the strong coupling superconductivity takes place in
underdoped cuprates and the pseudogap phenomena are its precursor.

We considered the T-Matrix approximation which is the dominant scattering
vertex in the vicinity of the superconducting critical point. In this
analysis we didn't considered the vertex corrections. Anyway our theory
appropriately explains the pseudogap phenomena in high-T$_c$ cuprates. We
did not take care into account the antiferromagnetic spin fluctuations which
are important for explaining other properties of cuprates such as the NMR
spin lattice relaxation $1/T_1$.

Based on this scenario we studied charge dynamics, in-plane and $c$-axis
resistivity and thermopower and compared the results with the experimental
results. As can be seen from previous section our theory fit well the
experimental results from $YBCO$ class of layered superconductors. For the
calculation of in-plane charge dynamics we obtained a spectrum dominated by
a Drude term, while for $c$-axis dynamics we found an almost frequency
independent behavior. For the $c$-axis resistivity we found, due to the
opening of the pseudogap in the density of states, a semiconductor-like
behavior. For the temperature dependence of thermopower we found a linearly
dependence at high temperatures with a negative slope. The negative slope
depends weakly on the doping level.

\section{Acknowledgments}

The research was supported by the CNCSIS contract no. 3447/1999.

\newpage\ 

\begin{center}
\textbf{Figure Captions}
\end{center}

Figure 1

Spectral function in the lowest order approximation at $T=0.6t$ in the $%
\left( 0,\pi \right) -\left( \pi ,\pi \right) $ direction at the doping
concentration $\delta =0.1$ and with the coupling strength $V=-5t$

Figure 2

Spectral function in the lowest order approximation at $T=0.6t$ in the $%
\left( 0,0\right) -\left( \pi ,\pi \right) $ direction at the doping
concentration $\delta =0.1$ and with the coupling strength $V=-5t$

Figure 3

Spectral function in the self consistent approximation at $T=0.6t$ in the $%
\left( 0,\pi \right) -\left( \pi ,\pi \right) $ direction at the doping
concentration $\delta =0.1$ and with the coupling strength $V=-5t$

Figure 4

Spectral function in the self consistent approximation at $T=0.6t$ in the $%
\left( 0,0\right) -\left( \pi ,\pi \right) $ direction at the doping
concentration $\delta =0.1$ and with the coupling strength $V=-5t$

Figure 5

Phase diagram of the model at hole concentration $\delta =0.1$

Figure 6 a

Density of states for temperature $T=0.35t$ and for different coupling
strength. The solid line correspond to $V=-t$, the dotted line to $V=-3t$
and the dashed line to $V=-6t$. The hole concentration is $\delta =0.1$.

Figure 6 b

Density of states for different temperatures and for coupling strength $%
V=-3t $. The solid line correspond to $T=3.5t$ and the dashed line to $%
T=0.35t$ at hole concentration is $\delta =0.1$.

Figure 7

In-plane optical conductivity at $T=0.4t$ (solid line) and $T=3.4t$ (dashed
line) at the doping concentration $\delta =0.1$ and coupling strength $%
V=-5.5t$

Figure 8

$c$-axis optical conductivity at different temperatures. $T=0.6t$ (solid
line) $T=1.0t$ (dash line) $T=1.6t$ (dot line) in the limit of small
frequencies at the coupling strength $V=-5.5t$ and doping level $\delta =0.1$

Figure 9

Temperature dependence of the in-plane resistivity at different densities, $%
\delta =0.1$ (solid line) ($T_C=0.21t$), $\delta =0.15$ ( dashed line) ($%
T_C=0.3t$) and $\delta =0.2$ (dashed-dotted line) ($T_C=0.36t$), at the same
coupling strength $V=-5.5t$. The arrows indicate the $T^{*}$ .

Figure 10

Temperature dependence of the $c$-axis resistivity at different densities, $%
\delta =0.1$ (solid line) ($T_C=0.21t$), $\delta =0.15$ ( dashed line) ($%
T_C=0.3t$) and $\delta =0.2$ (dotted line) ($T_C=0.36t$), at the same
coupling strength $V=-5.5t.$

Figure 11

Temperature dependence of thermopower at different densities, $\delta =0.1$
(solid line) ($T_C=0.21t$), $\delta =0.15$ ( dashed line) ($T_C=0.3t$) and $%
\delta =0.2$ (dotted line) ($T_C=0.36t$), at the same coupling strength $%
V=-5.5t.$

Figure 12

Termopower as function of the doping concentration $\delta $ at the coupling
strength $V=-5.5t$ and $T=2t$

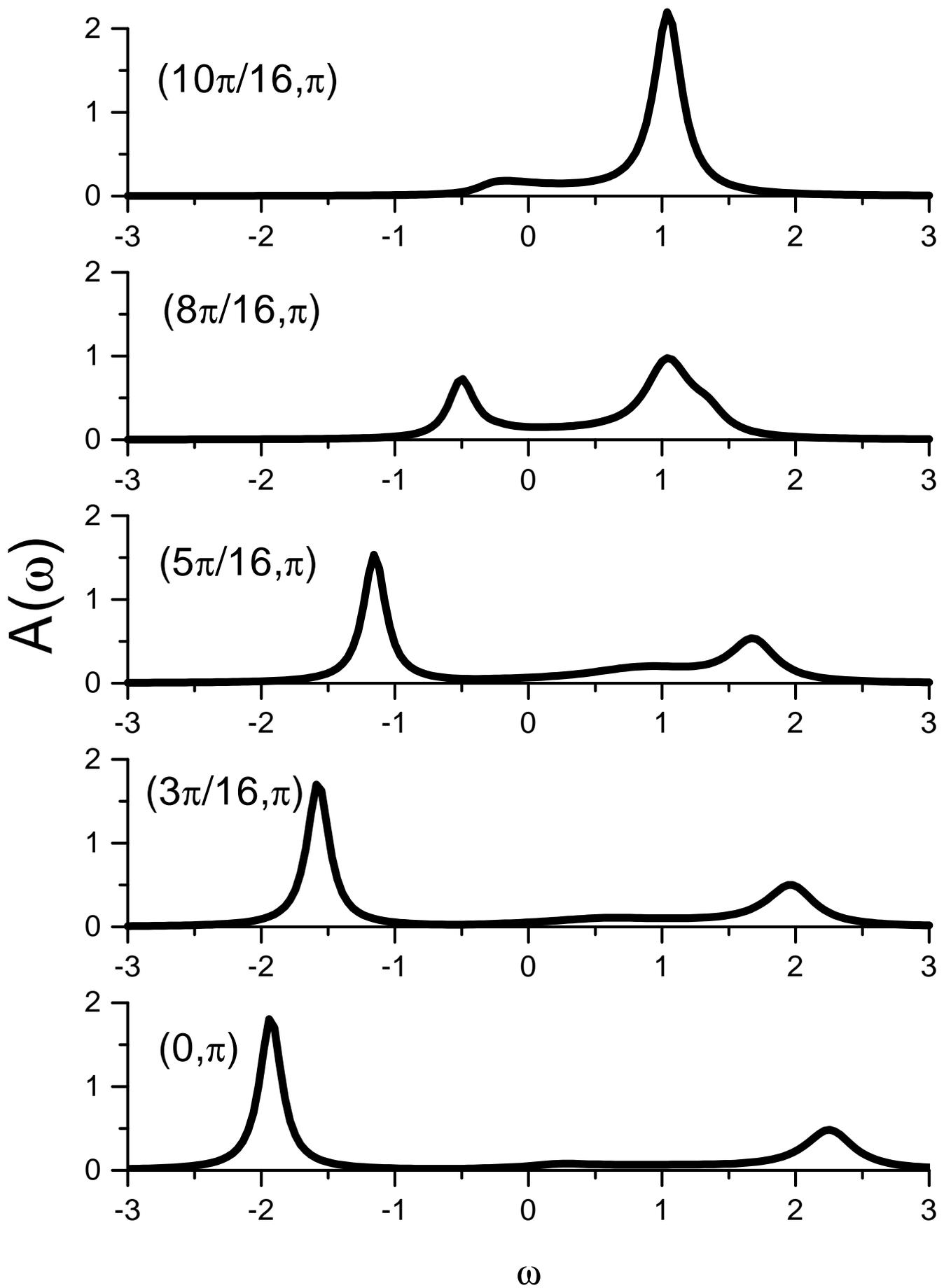
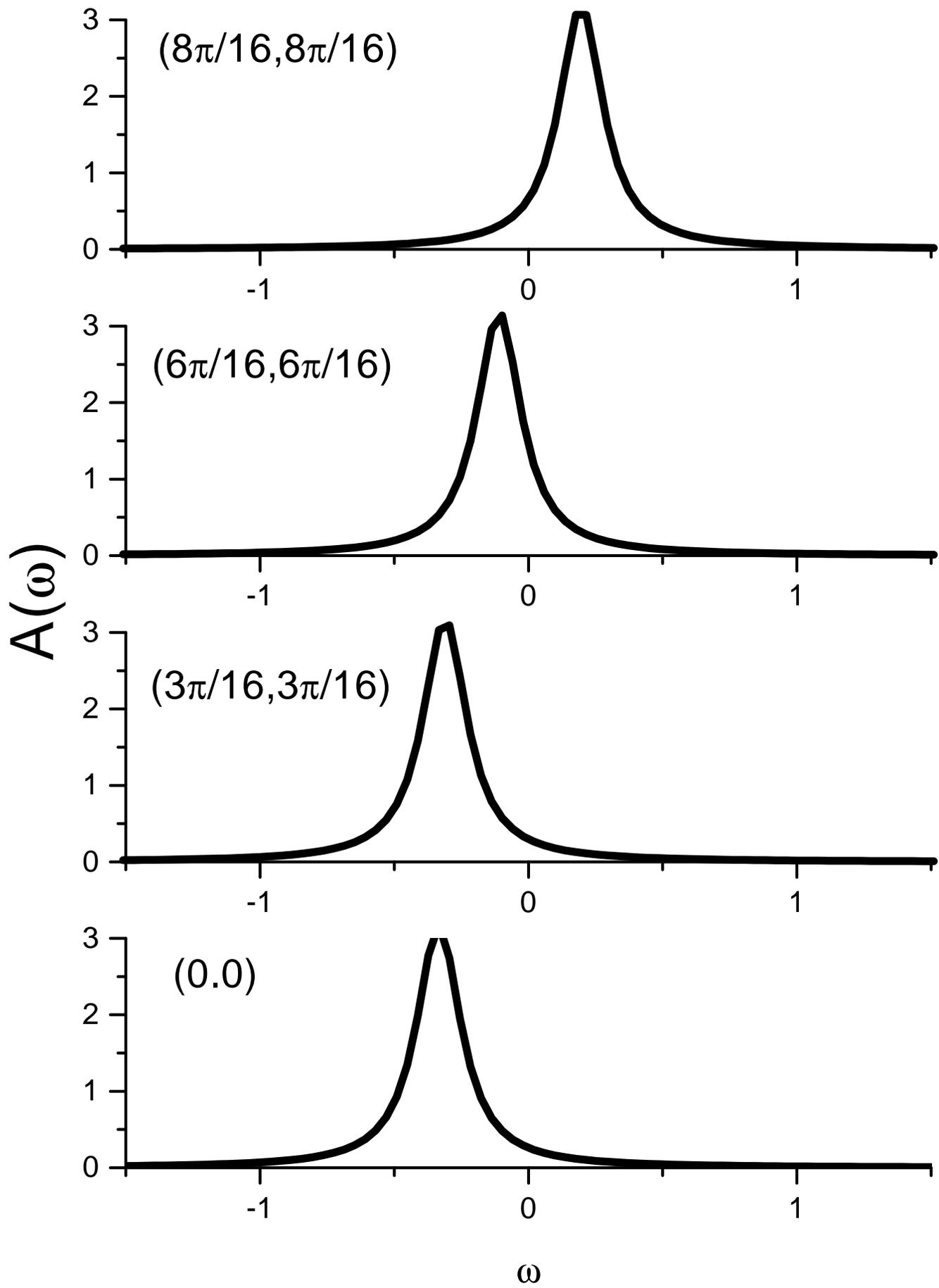
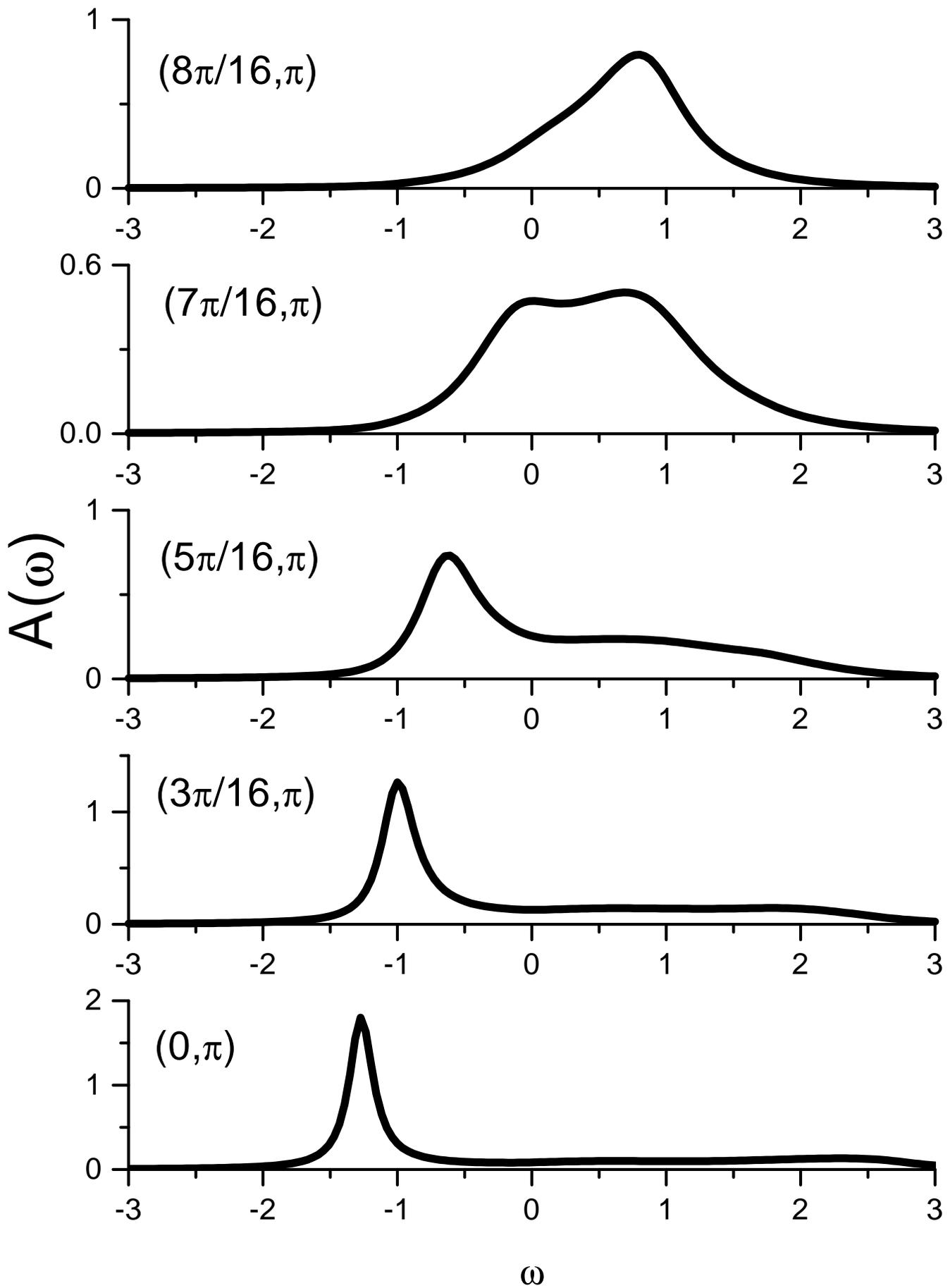
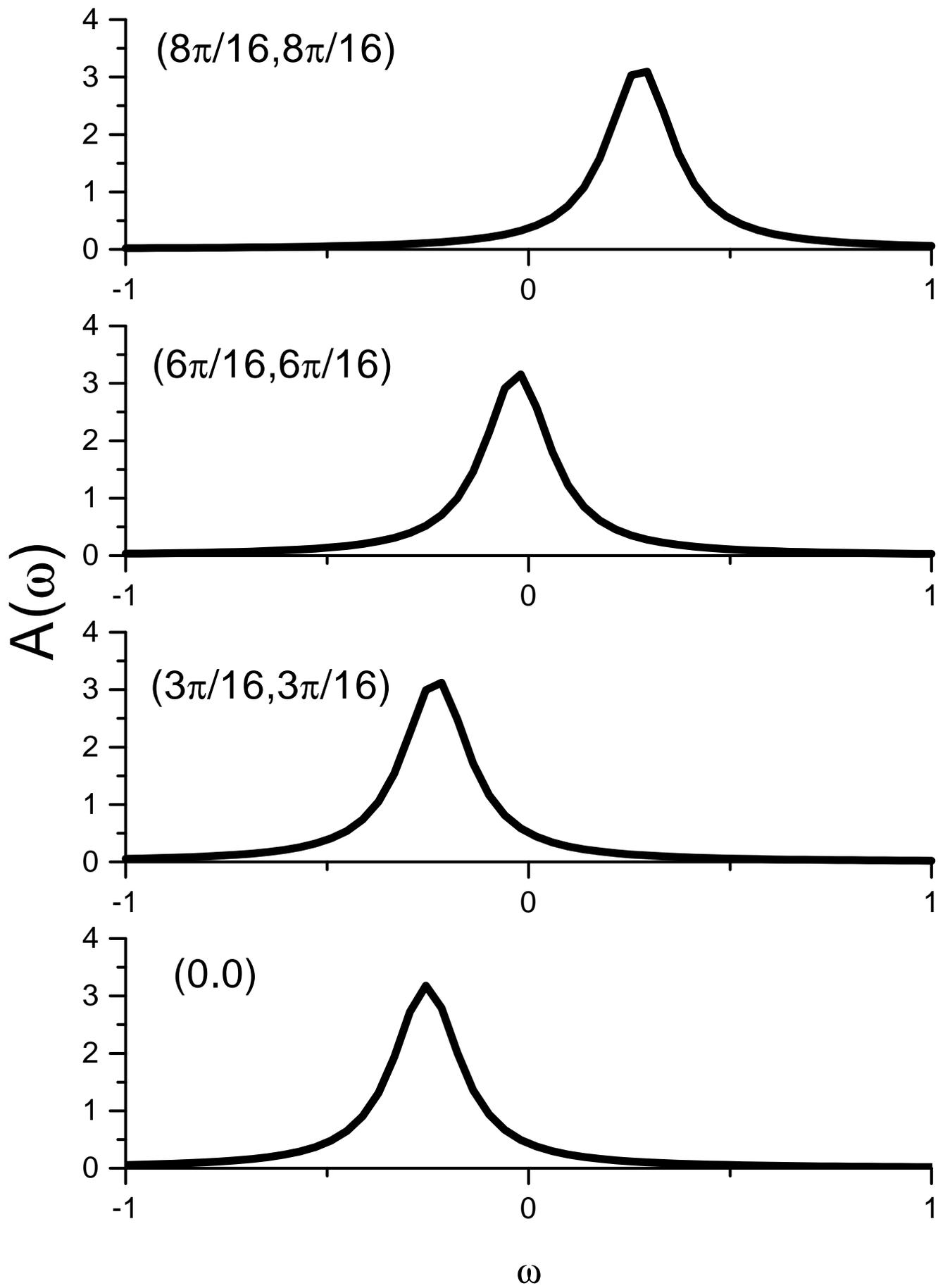
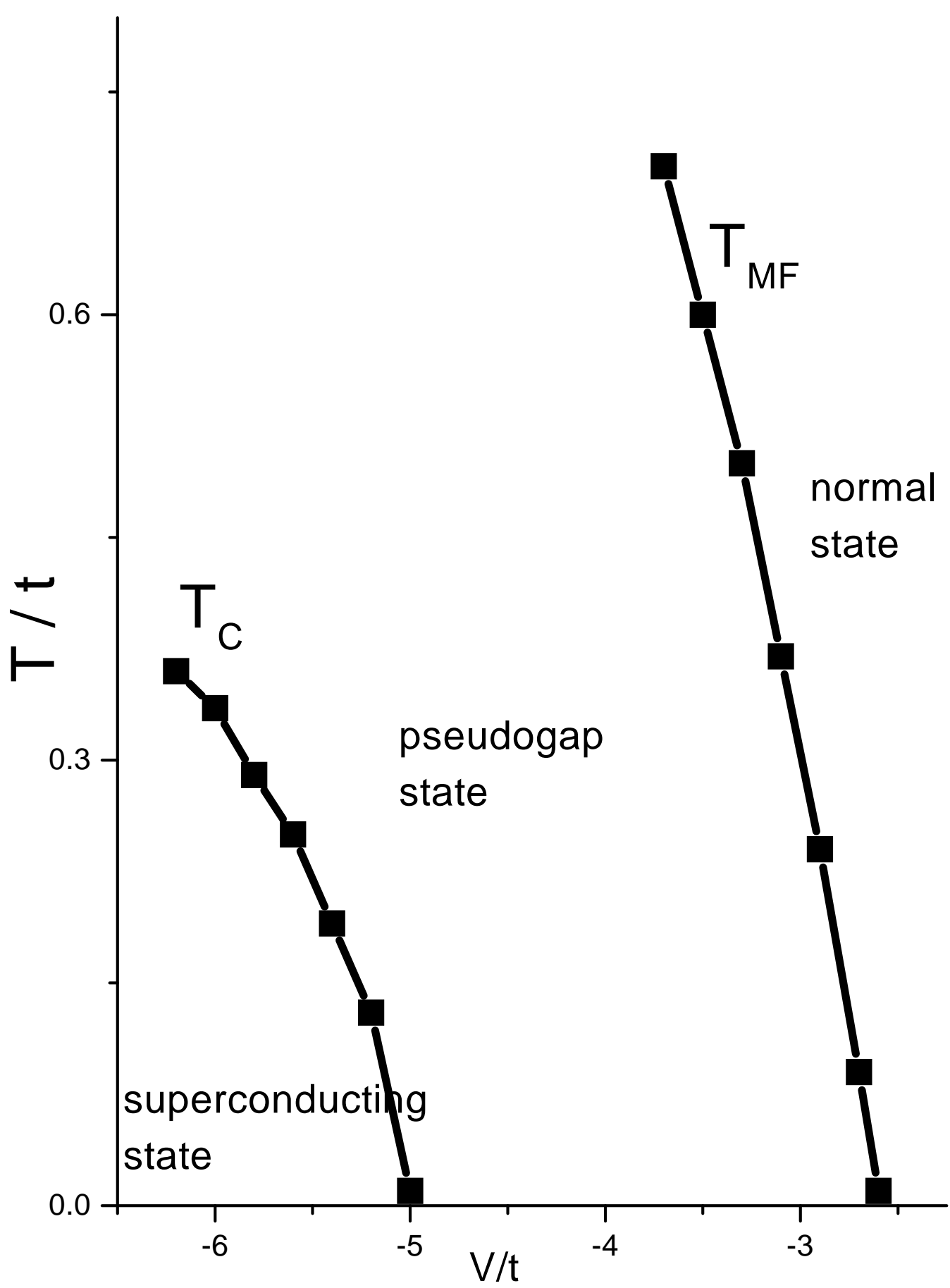
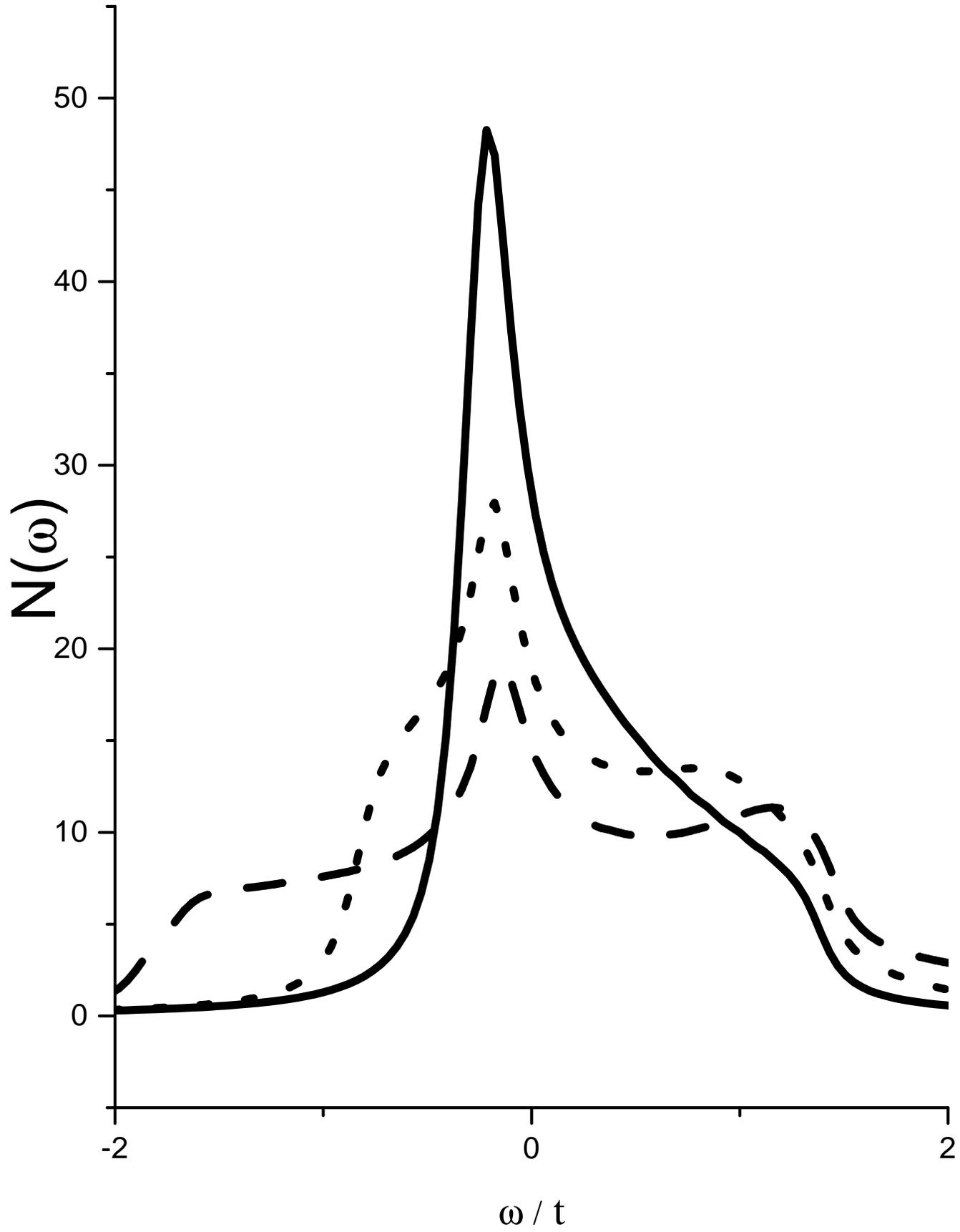
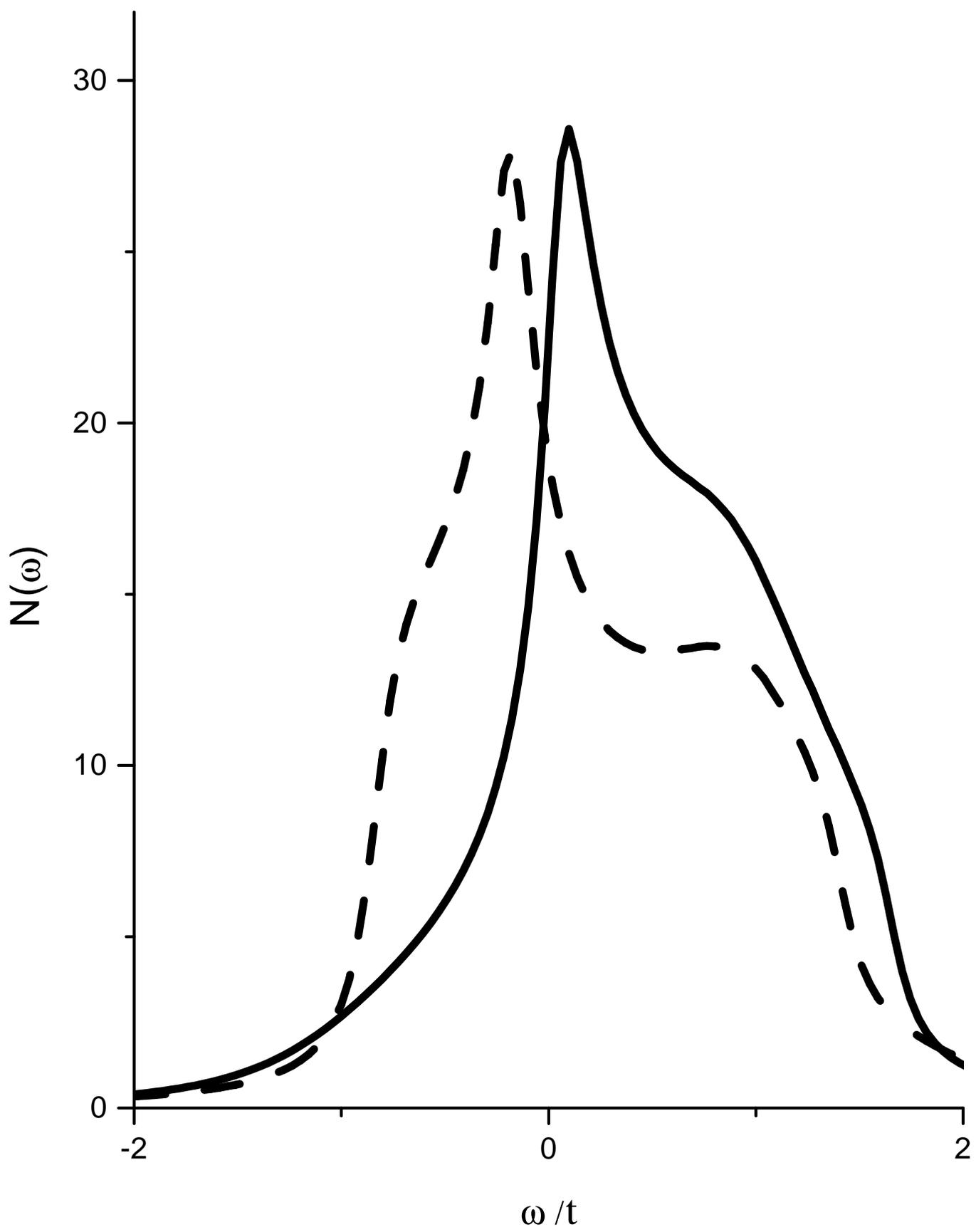
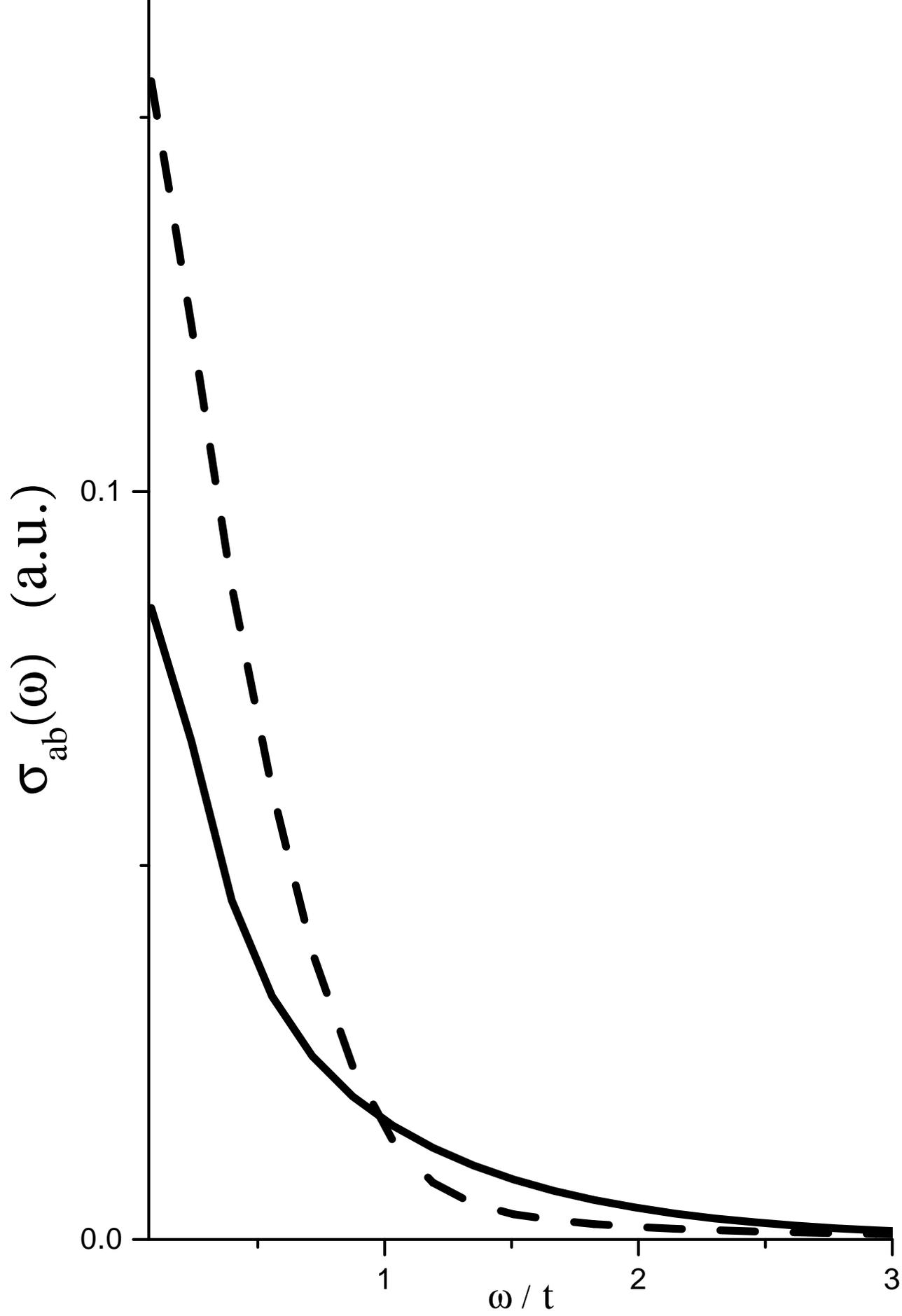
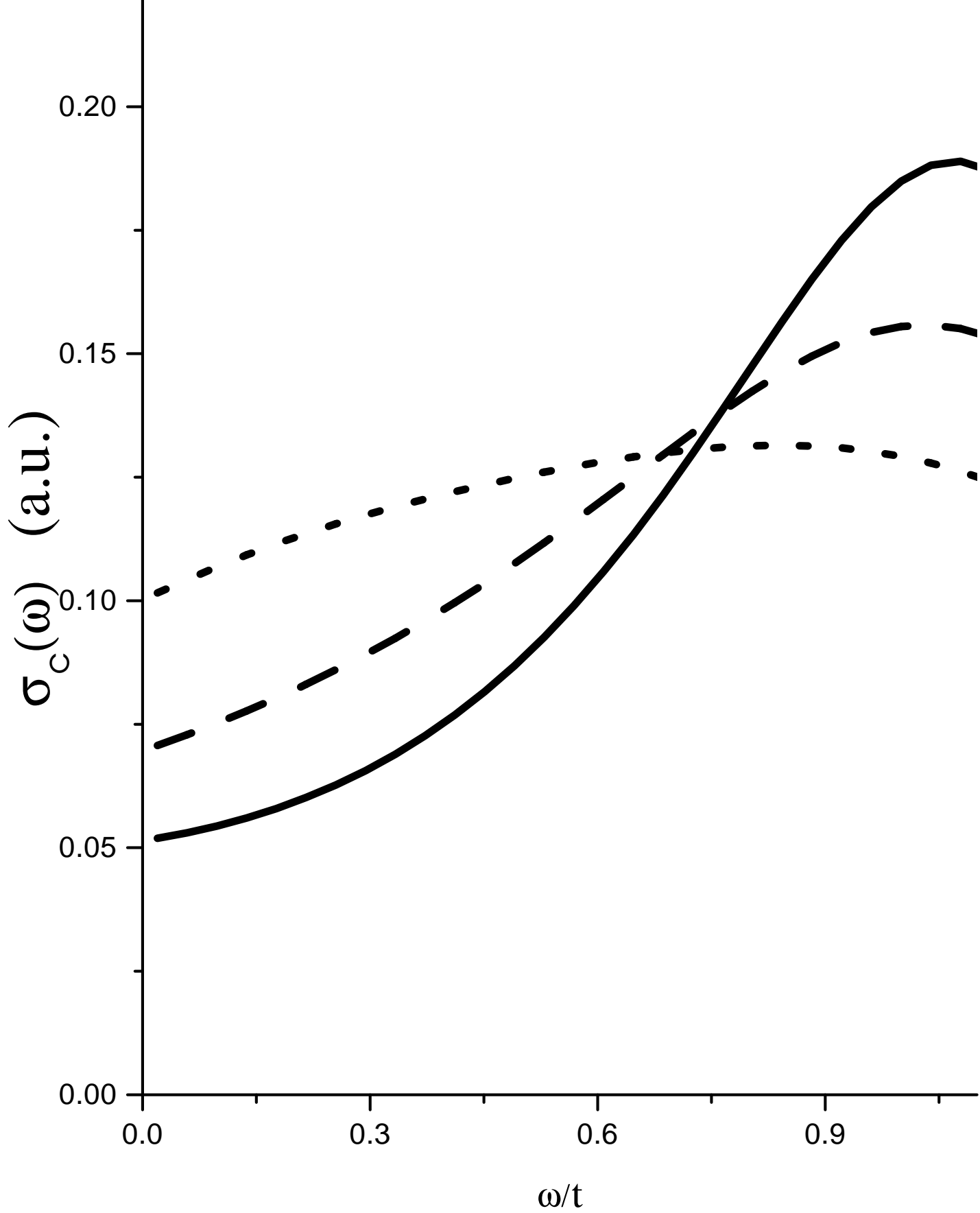
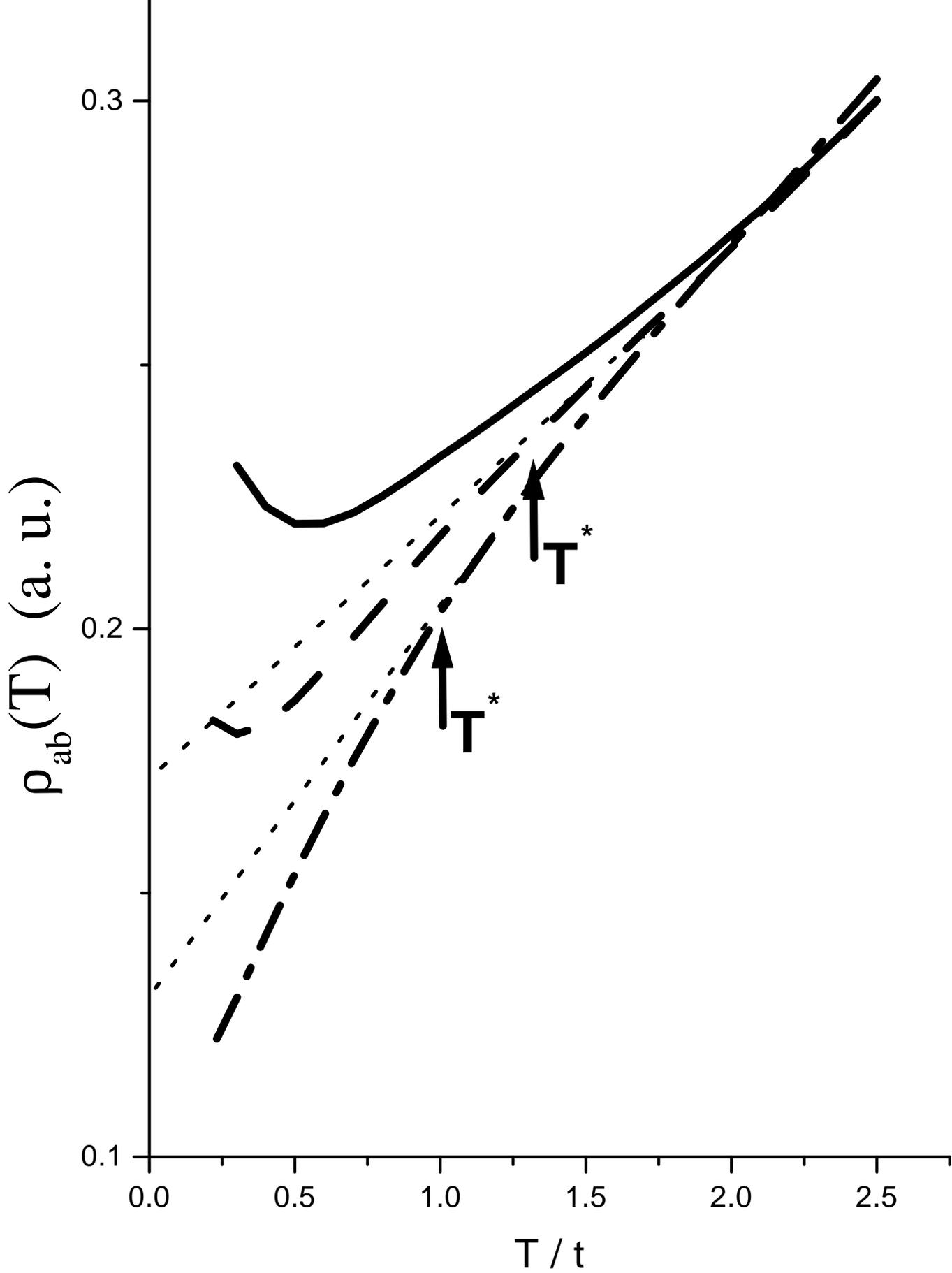
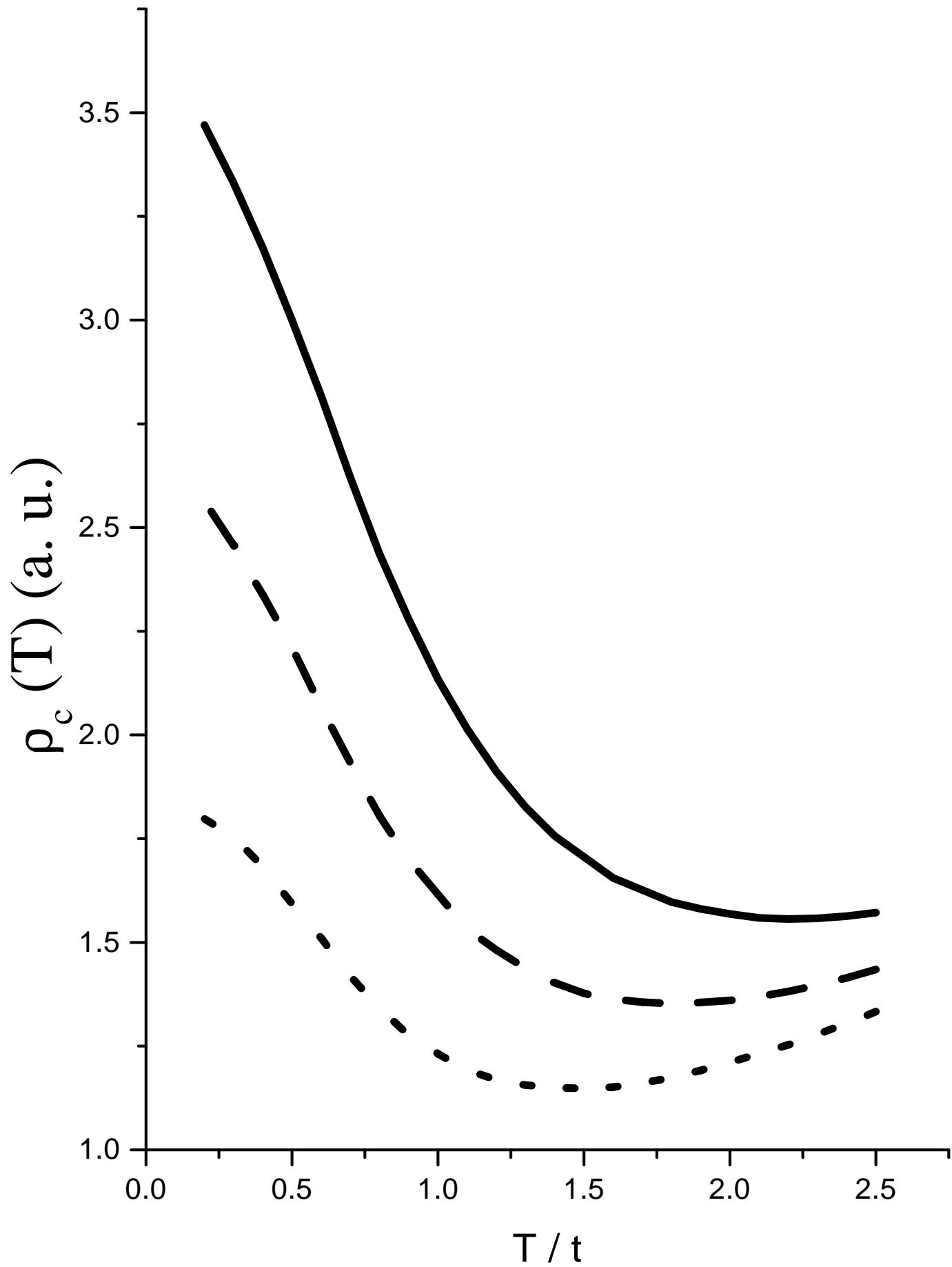
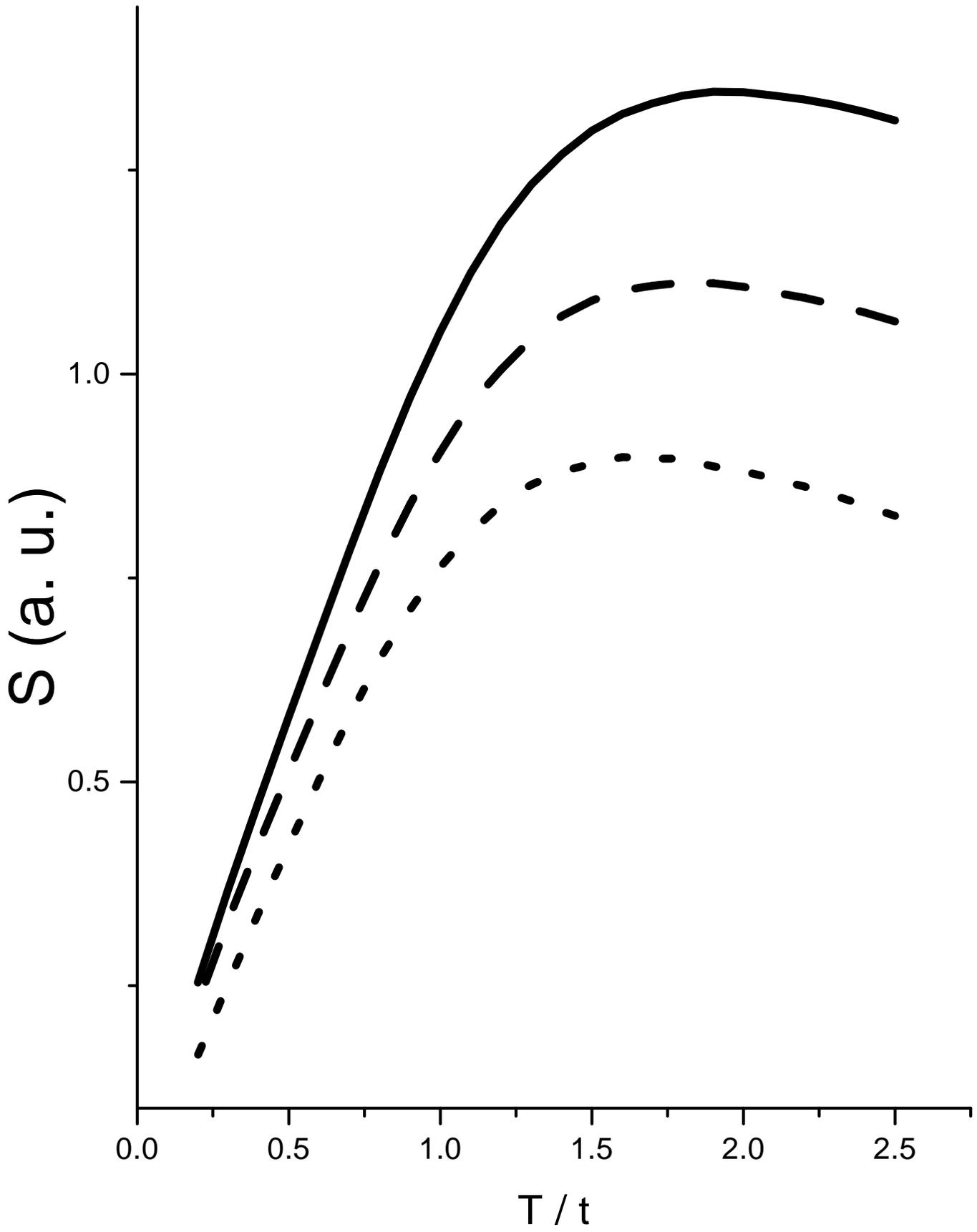
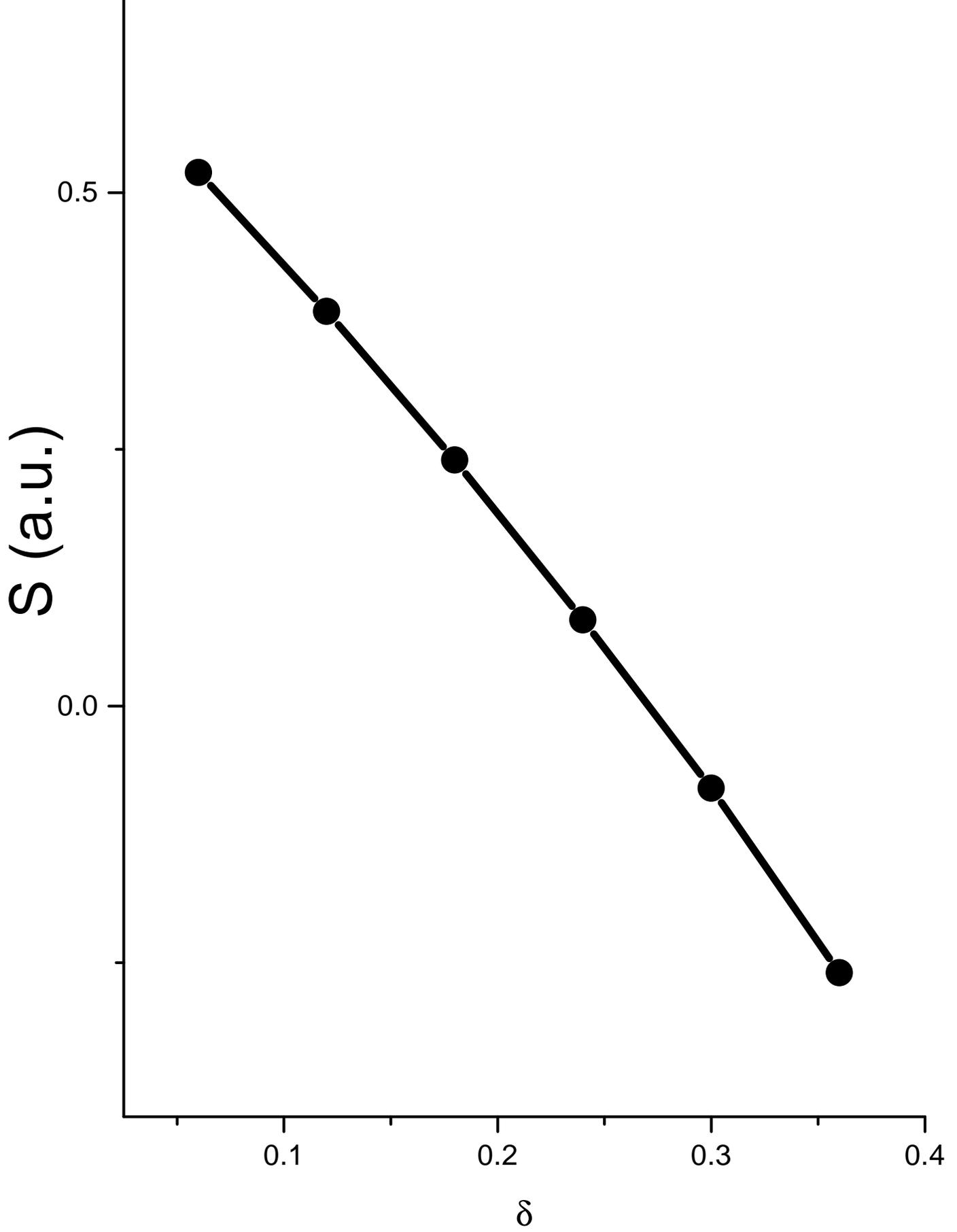
\end{document}